\documentstyle[aps,preprint]{revtex}
\tightenlines
\begin{document}
\draft
\title{\bf The Hubbard Model, Spinless Fermions, and Ising Spins} 
\author{K. Ziegler\\
Institut f\"ur Physik, Universit\"at Augsburg\\
D-86135 Augsburg, Germany\\
%e-mail: ziegler@physik.uni-augsburg.de
}
\maketitle
\begin{abstract}
The Hubbard model is used to study an electronic system at half filling. 
Starting from a functional integral representation the spin-up Grassmann
field is integrated out. It is shown that the resulting spinless fermion
theory has an instantaneous cluster interaction, and that the spinless
fermions
are coupled to thermally fluctuating Ising spins. The coupling parameter 
of the spinless fermion interaction is a product of the Hubbard interaction
and the hopping rate. As an example the strongly metallic as well as the
strongly insulating regime are investigated in terms of the effective Ising
statistics.
\end{abstract}

\section{Introduction}

\noindent
The Hubbard model was originally constructed to describe 
a metal-insulator transition for spin-dependent fermions
in a simple way \cite{hubbard,fradkin,rasetti,fulde,gebhard}. This transition
reflects the competition between potential (static) energy and kinetic
energy. The model is defined
on a lattice, where the potential energy consists of a chemical potential
and an on-site repulsion of fermions with opposite spin. The kinetic
energy is given by a nearest neighbor hopping. It turned out from
a number of calculations that this model has a rich structure because
of the complicated interplay of charge and spin degrees of freedom.
For instance, mean-field calculations for a magnetic order parameter
indicate para-, ferro- and antiferromagnetic states for
the half-filled system \cite{fradkin}. Thus, the magnetic properties of the
model became a central subject of investigations in solid state physics. 

The metal-insulator transition
was discussed originally by Hubbard using self-consistent approximations
\cite{hubbard}, later in terms of a variational approach
\cite{gutzwiller,brinkman},
and in the limit of an infinite dimensional lattice
\cite{metzner,vollhardt,gebhard}.
Very interesting investigations were obtained from computer simulations
which indicate an insulating phase at half filling for sufficiently strong
fermion interaction \cite{assaad,staudt}.
However, the detailed mechanism and the properties of the
transition are not entirely clear.

To study the metal-insulator transition
one can, in principle, start either from the metallic or from the
insulating side. As the simplest approximations one could use non-interacting
fermions on the metallic side or the local limit on the
insulating side, where the hopping rate is zero. Unfortunately, neither
of these starting points is very useful in order to understand the
interacting Hubbard model: Non-interacting fermions are unstable against
an arbitrarily weak interaction \cite{fradkin}, and the local limit is
completely degenerate with respect to the spin. Therefore, an arbitrarily
weak hopping rate would lift the degeneracy leading to a new state
which might be magnetically ordered \cite{tasaki}. The basic idea of the
present work is to start from the extreme insulating as well as from
the extreme metallic state at low temperatures and to construct a
perturbation theory without analyzing its magnetic order. The latter
is a restriction which simplifies the calculations significantly because
the spin degree of freedom can be ignored.

In this work a grand canonical ensemble is considered, where on average
one fermion per site (half-filled system) is assumed. The
non-interacting fermions as well as the static fermions (i.e. fermions
without a hopping term) have a $2^M$-degeneracy
($M$ is the number of lattice sites) because each site can accomodate a
fermion with spin-up or one with spin-down. Consequently, a perturbation
theory around one of these limiting states is plagued by the
degeneracies. For instance, a perturbation around the static
state is a spontaneous hop of a fermion from any
site to its nearest neighbor site. As a consequence, the fermion
spontaneously creates a doubly occupied site and an empty site. 
The doubly occupied site may decay after some time again into
two singly occupied sites. The resulting state is two-fold
degenerate because of the possible two spin orientations.
The unperturbed state can be an antiferromagnetic
(N\'eel) state. A hopping process at a time $t_1$ can exchange two neighboring
fermions which leads to two pairs of neighboring fermions with parallel spins.
At time $t_2$ the inverse hopping process can re-create the original
antiferromagnetic state.  Therefore, the two hopping processes are not
independent.
Moreover, the intermediate state between time $t_1$ and $t_2$ has the same
energy as the antiferromagnetic state. This implies a constant interaction
in time. Consequently, the linked cluster theorem cannot be
applied, since it works only for independent clusters or clusters which
interact with a decaying interaction \cite{glimm,brydges}. The central point
of the present work is a concept which deals with this degeneracy.

In order to control this exponential degeneracy it is natural to
eliminate one spin orientation. This can be achieved formally by
integrating out one of the spin orientation in the functional
integral representation of the Hubbard model. The result of this
operation reveals an important structure of the effective spinless
fermion model which is formally an expansion of the model in
terms of the degeneracy: the expansion terms are not degenerate
and the perturbation expansion can be applied independently to
each of them. It turns out that the expansion is equivalent to
the summation over the $2^M$ states of thermal Ising spins which
are coupled to the spinless fermions. After an approximation which
is applicable for the strongly metallic and the strongly insulating
regime of the Hubbard model, the fermionic degrees of freedom can
be integrated out. Thus, the physics is described by the Ising
spins: The strongly metallic regime is characterized by a
ferromagnetic Ising structure in which the fermions can freely
move at low temperatures. On the contrary, the strongly insulating
regime is characterized by an antiferromagnetic Ising structure
which creates a gap for the fermions, in formal analogy to a
Peierls instability. 

The article is organized as follows:
In Sect. II the Hubbard model is defined in a coherent state representation
for a grand canonical ensemble of fermions.
The static limit (no hopping) of the Hubbard is briefly discussed in
Sect. III.
Then in Sect. IV the integration over the spin-up component of the
model is performed. The resulting model of spinless fermions, which has
a complicated but instantaneous cluster interaction, is analyzed in 
Sect. IV.A. In Sect. V the Ising spin representation of the spinless fermion
model is introduced and discussed. Finally, in Sect. V.A the
weak-coupling limit and in Sect. V.B the weak hopping limit are
studied. Appendices A, B, and C give details of the calculations.  

\section{The Hubbard Model}

The Hubbard model describes fermions with spin $\sigma=\downarrow,
\uparrow$ on a $d$-dimensional lattice $\lambda$. It is defined by the
Hamiltonian \cite{fradkin,gebhard}
\[
H[c_\sigma^\dagger(r),c_\sigma(r)]
= -{\bar t}\sum_{\langle r,r'\rangle ,\sigma}c_\sigma^\dagger(r)
c_\sigma(r')  + \sum_r\Big[\mu\sum_\sigma 
c_\sigma^\dagger(r)c_\sigma(r)+Uc^\dagger_\uparrow(r)c_\uparrow(r)
c^\dagger_\downarrow(r)c_\downarrow(r)\Big],
\]
where $c_\sigma^\dagger(r)$, $c_\sigma(r)$ are fermion creation and 
annihilation operators, respectively.
${\bar t}\ge0$ is the hopping rate. $\langle r,r'\rangle$ means pairs of
nearest neighbor sites on the lattice. $\mu$ is the chemical potential.

Using this Hamiltonian a grand canonical ensemble of fermions at the
inverse temperature $\beta$ can be
defined by the partition function, given in terms of a functional
integral (coherent state representation) on a Grassmann algebra
\cite{negele}. For the latter the integration over
a complex Grassmann field $(\Psi_\sigma(r,t),{\bar\Psi}_\sigma(r,t))$
is given as a linear mapping from a Grassmann algebra to the complex
numbers. At a space-time point $(r,t)$ we have for integers $k,l\ge0$
\[
\int [{\bar\Psi}_\sigma(r,t)]^k[\Psi_\sigma(r,t)]^l
d\Psi_\sigma(r,t)d{\bar\Psi}_\sigma(r,t)=\delta_{k,1}\delta_{l,1}. 
\]
The partition function of the grand canonical ensemble of fermions then
reads
\[
Z=\int\exp(-S){\cal D}[\Psi,{\bar\Psi}]
\]
with the action
\begin{equation}
S=i\Delta\sum_{r,t}{1\over i\Delta}{\bar\Psi}_\sigma(r,t)
[\Psi_\sigma(r,t)-\Psi_\sigma(r,t-\Delta)]
+i\Delta\sum_t {1\over\hbar}
H[{\bar\Psi}_\sigma(r,t),\Psi_\sigma(r,t-\Delta)]
\label{action1}
\end{equation}
and the product measure
\[
{\cal D}[\Psi_\uparrow,\Psi_\downarrow]=\prod_{r,t,\sigma}
d\Psi_\sigma(r,t)d{\bar\Psi}_\sigma(r,t).
\]
The discrete time is used with $t=\Delta, 2\Delta,...,\beta$.
${\bar\Psi}_\sigma(r,t)$ and $\Psi_\sigma(r,t)$ are independent
Grassmann fields which satisfy antiperiodic boundary conditions in
time $\Psi_\sigma(r,\beta+\Delta)=-\Psi_\sigma(r,\Delta)$
and ${\bar\Psi}_\sigma(r,\beta+\Delta)=-{\bar\Psi}_\sigma(r,\Delta)$.
For the subsequent calculations it is convenient to rename 
$\Psi_\sigma(r,t)\to\Psi_\sigma(r,t+\Delta)$ because then the 
Grassmann field appears with the same time in the Hamiltonian of the
action (\ref{action1}).

\section{The Local Limit}

Neglecting the hopping term in the Hamiltonian (i.e., for ${\bar t}=0$), 
the integration in the partition function factorizes in space, and 
the corresponding expression can be evalueated as
\begin{equation}
Z=\int\prod_{r,t}\exp[-S_d(r,t)]{\cal D}[\Psi_\uparrow,\Psi_\downarrow]
=\prod_r\int \prod_t\exp[-S_d(r,t)]{\cal D}[\Psi_\uparrow,\Psi_\downarrow]
=Z_1^M,
\label{local}
\end{equation}
where
\[
S_d=\sum_{t,r}\Big\{{\bar\Psi}_\sigma(r,t)\Psi_\sigma(r,t+\Delta)
-{\bar\mu}{\bar\Psi}_\sigma(r,t)\Psi_\sigma(r,t)
+i\Delta U{\bar\Psi}_\uparrow(r,t)\Psi_\uparrow(r,t)
{\bar\Psi}_\downarrow(r,t)\Psi_\downarrow(r,t)\Big\}
\]
with $\hbar=1$ and ${\bar\mu}=1-i\Delta\mu$. $Z_1$ is the partition
function of the
Hubbard model with one lattice site (static, local or atomic limit):
\[
Z_1=\int\prod_t \exp[-S_d(r,t)]
d\mu(r,t)=1+2{\bar\mu}^{\beta/\Delta}
+[{\bar\mu}^2-i\Delta U]^{\beta/\Delta}.
\]
Using the new parameters $\mu'=i\Delta\mu$, $U'=i\Delta U$, and 
$\beta'=\beta/\Delta$, we can define the following weights,
depending on the number of particles per site
\begin{eqnarray}
w_0 & = & 1/Z_1
\nonumber\\
w_1 & = & 2(1-\mu')^{\beta'}/Z_1
\nonumber\\
w_2 & = & [(1-\mu')^2-U']^{\beta'}/Z_1.
\nonumber
\end{eqnarray}
In the temperature formalism, where the time $t$ is replaced by the
imaginary time through a Wick rotation, the weights $w_0$, $w_1$ 
and $w_2$ are statistical weights. Then the average number of
particles per site is
\[
n=w_1+2w_2.
\]
At zero temperature ($\beta'\to\infty$) this gives
\[
n=\cases{
0&if $0<\mu'<1+\sqrt{1+U'}$\cr
1&if $\mu'<0$ and $\mu'(\mu'-1)<U'$\cr
2& otherwise \cr
}.
\]
The groundstate of the expansion around the local limit is degenerated
because each singly occupied site can accomodate a fermion with either
spin $\uparrow$ or with spin $\downarrow$. This degeneracy must be handled
with care. In particular, to obtain a unique hopping expansion
one has to separate the
degenerate contributions. Therefore, one cannot directly work with
the hopping term as a perturbation but have to set up a perturbation theory 
which works with non-degenerate groundstates. This means that one has to
divide the degenerate groundstate such that the perturbations remain in 
their corresponding non-degenerate groundstates. 

%%%%%%%%%%%%%%\section{Non-degenerate Perturbation Theory}  

\section{Integration over the spin-up field $\Psi_\uparrow$}

The action $S$ can be divided into three pieces as
\[
S=S_\uparrow+S_\downarrow+S_I
\]
with
\[
S_\sigma=\sum_t\Big\{\sum_r
\big[{\bar\Psi}_\sigma(r,t)\Psi_\sigma(r,t+\Delta)
-{\bar\mu}{\bar\Psi}_\sigma(r,t)\Psi_\sigma(r,t)\big]
-\tau\sum_{\langle r,r'\rangle}{\bar\Psi}_\sigma(r,t)
\Psi_\sigma(r',t)\Big\}
\]
for $\sigma={\uparrow,\downarrow}$ with $\tau=i\Delta{\bar t}$. The
interaction between the two spin orientations is given by
\[
S_I=U'\sum_{r,t}
{\bar\Psi}_\uparrow(r,t)\Psi_\uparrow(r,t)
{\bar\Psi}_\downarrow(r,t)\Psi_\downarrow(r,t).
\]
Now it is possible to integrate out the spin-up field $\Psi_\uparrow$,
since the field $\Psi_\uparrow$ appears in $S$ only as a quadratic form.
The integration over this Grassmann field gives a determinant
\begin{equation}
\int e^{-S_\uparrow-S_I}\prod_{r,t}
d\Psi_\uparrow(r,t)d{\bar\Psi}_\uparrow(r,t)
=det\big[-\partial_t+{\bar\mu}+{\hat t}
-U'{\bar\Psi}_\downarrow\Psi_\downarrow\big],
\label{det}
\end{equation}
where $\partial_t$ is the time-shift operator
\[
\partial_t\Psi(r,t)=\cases{
\Psi(r,t+\Delta) & $\Delta\le t < \beta$ \cr
-\Psi(r,\Delta) & $t=\beta$ \cr
}.
\]
The last equation is due to the antiperiodic boundary condition of the
Grassmann field. This definition gives
\[
(\partial_t)^{-1}=\partial_t^T,\ \ \ det(\partial_t)=1.
\]
We assume that the number of time slices $\beta'$ is even such that
$det(-\partial_t)=det(\partial_t)=1$. The matrix ${\hat t}_{r,r'}=\tau$
if $r,r'$ are nearest neighbors and zero otherwise.
Expressions in the determinant which do not have a specified matrix
structure are implicitly multiplied by the corresponding
unit matrix. For instance, ${\bar\mu}$ is multiplied by the space-time
unit matrix whereas ${\hat t}$ is multiplied by the time-like unit
matrix. 

In the following subsets of the space-time lattice
$\Lambda=\lambda\otimes\{\Delta, 2\Delta,...,\beta\}$ will be considered.
For a subset $\Lambda_k\subset\Lambda$ we define the determinant of the
the projected matrix $P_k AP_k$ as
\[
det_{\Lambda_k}A\equiv det_{\Lambda_k}(P_kAP_k),
\]
where $P_k$ is the projector onto $\Lambda_k$. 
\\

\subsection{Effective Cluster Action of Spinless Fermions}

The partition function is now a functional integral of the
spin-down Grassmann field
\[
Z=\int e^{-S_\downarrow}det\big[-\partial_t+{\bar\mu}+{\hat t}
-U'{\bar\Psi}_\downarrow\Psi_\downarrow\big]
{\cal D}[\Psi_\downarrow].
\]
Formally, the determinant could be expressed as part of
the action by using the identity $det A = \exp[Tr(\log A)]$. However,
this would be too naive because the term 
\[
Tr\Big[\log\Big( -\partial_t+{\bar\mu}+{\hat t}
-U'{\bar\Psi}_\downarrow\Psi_\downarrow\Big)\Big]
\]
has a complicated interaction of the Grassmann field in space
{\it and} time. Moreover, at least for ${\bar t}=0$ the interaction has
a long-range part in time which reflects the degeneracy of the unperturbed
system. Fortunately, there is a way to avoid these difficulties:
As shown in Appendix A, the determinant can be expanded in terms of the 
partitions $\Lambda_k\subseteq\Lambda$ of the space-time lattice
$\Lambda$ as
\[
det\big[-\partial_t+{\bar\mu}+{\hat t}
-U'{\bar\Psi}_\downarrow\Psi_\downarrow\big]
=\sum_{\Lambda_k\subseteq\Lambda}
det_{\Lambda_k}[-({\bar\mu}+{\hat t})\partial_t^T]
\exp\Big[
Tr_{\Lambda_k}\log\Big(
{\bf 1}-({\hat t}+{\bar\mu})^{-1}
U'{\bar\Psi}_\downarrow\Psi_\downarrow\Big)
\Big].
\]
The partitions include the empty set which gives $det_{\emptyset}A=1$.
This expansion is the most important step for the treatment of the
Hubbard model in this work. The first consequence is that the partition
function $Z$ is now given by an expansion in terms of $\Lambda_k$ as
$Z=\sum_{\Lambda_k}Z_{\Lambda_k}$ with
\begin{equation}
Z_{\Lambda_k}=
det_{\Lambda_k}[-({\bar\mu}+{\hat t})\partial_t^T]
\int\exp(-S_{\Lambda_k}){\cal D}[\Psi_\downarrow]
\label{a}
\end{equation}
and the action
\[
S_{\Lambda_k}={\bar\Psi}_\downarrow
\cdot(\partial_t-{\bar\mu}-{\hat t})\Psi_\downarrow
-Tr_{\Lambda_k}\log\Big(
{\bf 1}-({\hat t}+{\bar\mu})^{-1}
U'{\bar\Psi}_\downarrow\Psi_\downarrow\Big).
\]
The second term of $S_{\Lambda_k}$ can be expanded in powers of
the Grassmann field. This yields an {\it instantaneous} cluster
interaction
\[
S_i=\sum_{l\ge1}{{U'}^l\over l}\sum_{(r,t)\in\Lambda_k}
\sum_{r_1,...,r_{l-1}}({\hat t}+{\bar\mu})^{-1}_{r,r_1}
{\bar\Psi}_\downarrow(r_1,t)\Psi_\downarrow(r_1,t)
...
({\hat t}+{\bar\mu})^{-1}_{r_{l-1},r}
{\bar\Psi}_\downarrow(r,t)\Psi_\downarrow(r,t).
\]
Due to the identity
$$
({\hat t}+{\bar\mu})^{-1}={1\over{\bar\mu}}-
{{\hat t}\over{\bar\mu}}({\hat t}+{\bar\mu})^{-1}
$$
the action $S_i$ can also be written with 
$u=U'({\hat t}+{\bar\mu})^{-1}_{rr}$ as
\[
S_i=u\sum_{(r,t)\in\Lambda_k}
{\bar\Psi}_\downarrow(r,t)\Psi_\downarrow(r,t)
\]
\[
+\sum_{l\ge2}{{U'}^l\over l}\sum_{(r,t)\in\Lambda_k}
\sum_{r_1,...,r_{l-1}}\Big({-{\hat t}\over{\bar\mu}}
({\hat t}+{\bar\mu})^{-1}\Big)_{r,r_1}
{\bar\Psi}_\downarrow(r_1,t)\Psi_\downarrow(r_1,t)
\]
\[ ...
\Big({-{\hat t}\over{\bar\mu}}({\hat t}+{\bar\mu})^{-1}
\Big)_{r_{l-1},r}
{\bar\Psi}_\downarrow(r,t)\Psi_\downarrow(r,t).
\]
This shows that the interaction terms with $l\ge2$ do not distinguish
between weak interaction ($U'\sim0$) and weak hopping (${\bar t}\sim0$)
because the expansion parameter is $U'{\bar t}$.

The non-interacting limit $U'=0$ as well as the local limit ${\hat t}=0$ 
can be checked immediately because of a vanishing interaction. The
former gives the product of two determinants (from both spin orientations)
\[
Z_0=det\big(-\partial_t+{\bar\mu}+{\hat t}\big)^2
\]
and the latter
\[
Z=Z_1^M,
\]
in agreement with Eq. (\ref{local}).

\section{Ising Spin Statistics}

It is interesting to notice that the determinant
$det_{\Lambda_k}[-({\bar\mu}+{\hat t})\partial_t^T]$
in $Z_{\Lambda_k}$ determines which partition $\Lambda_k$ contributes with
non-zero weight. Since $\partial_t$
is diagonal in space but off-diagonal in time, the determinant 
$det_{\Lambda_k}(\partial_t^T)$ is non-zero only if a site $r$ is in
$\Lambda_k$ at all times. Therefore, the contributing partitions are of the
form $\Lambda_k=\lambda_k\times\{\Delta,2\Delta,...,\beta\}$, where
$\lambda_k$ is a partition of the space lattice $\lambda$. The 
determinant then reads
\[
det_{\Lambda_k}[-({\bar\mu}+{\hat t})\partial_t^T]
=[det_{\lambda_k}({\bar\mu}+{\hat t})]^{\beta'}.
\]
Now it is convenient to introduce the projector $I_{\lambda_k}$ which
projects on $\lambda_k$
\[
(I_{\lambda_k})_r=\cases{
1 & if $r\in\lambda_k$ \cr
0 & if $r\notin\lambda_k$ \cr
}.
\]
Consequently, the determinant can be expressed as
\begin{equation}
[det_{\lambda_k}({\bar \mu}+{\hat t})]^{\beta'}
=\Big[det_\lambda({\bf 1}+ I_{\lambda_k}[{\bar \mu}+{\hat t}-{\bf 1}]
I_{\lambda_k})\Big]^{\beta'}.
\label{b}
\end{equation}
The interaction can also be written in terms of the projection
operator:
\[
S_i=\sum_{l\ge1}{{U'}^l\over l}\sum_t
\sum_{r,r_1,...,r_{l-1}}(I_{\lambda_k})_r
({\hat t}+{\bar\mu})^{-1}_{r,r_1}
{\bar\Psi}_\downarrow(r_1,t)\Psi_\downarrow(r_1,t)
...
({\hat t}+{\bar\mu})^{-1}_{r_{l-1},r}
{\bar\Psi}_\downarrow(r,t)\Psi_\downarrow(r,t)
\]
The partitions $\lambda_k$ enter $Z$ only through the projection operator
$I_{\lambda_k}$. Eqs. (\ref{a}) and (\ref{b}) imply
\[
Z=\sum_{\lambda_k}Z_{\lambda_k}=
\sum_{\lambda_k}
\Big[det_\lambda({\bf 1}+ I_{\lambda_k}[{\bar \mu}+{\hat t}-{\bf 1}]
I_{\lambda_k})\Big]^{\beta'}
\int\exp(-S_{\Lambda_k}){\cal D}[\Psi_\downarrow]
\]
with the action
\[
S_{\lambda_k}=
{\bar\Psi}_\downarrow\cdot(\partial_t-{\hat t}-{\bar\mu}
+uI_{\lambda_k})\Psi_\downarrow
+O((U'{\bar t}/{\bar\mu}^2)^2).
\]
The projection operator $I_{\lambda_k}$ can be expressed
as $(I(\{S_r\}))_{rr'}=(1+S_r)\delta_{rr'}/2$ with the Ising spin
\[
S_r=\cases{
1 & $r\in\lambda_k$ \cr
-1 & $r\notin\lambda_k$ \cr
}.
\]
The partitions $\lambda_k$ enter $Z$ only through the projection operator
$I_{\lambda_k}$.
Therefore, the sum over the randomly chosen partitions $\lambda_k$ of
the space lattice is equivalent to the sum over randomly chosen Ising
spins.
This is a thermal Ising model with complicated interaction spin
interaction
in a magnetic field. The latter is implied by the missing invariance
under a global transformation $S_r\to -S_r$.

Neglecting the interaction $S_i$, which is weak for the
metallic as well as for the insulating regime, one gets
\begin{equation}
Z\approx Z_I=\sum_{\{ S_r=\pm1\}}
\Big[det_\lambda({\bf 1}+ I[{\bar \mu}+{\hat t}-{\bf 1}]
I)\Big]^{\beta'}
det(-\partial_t+{\bar\mu}-uI+{\hat t}).
\label{partitionf}
\end{equation}
The partition function $Z_I$ can be used as a starting point for an
approximative treatment
of the Hubbard model. In the rest of this work the favoured spin
structures will be analyzed, i.e., the spin configurations with maximal
Boltzmann weight. Details of the calculations are given in App. B and C.

\subsection{Weak Interaction: $U'\approx0$}

A metallic behavior is expected in this regime. Since 
\[
[det_{\lambda_k}({\bar\mu}+{\hat t})]^{\beta'}
\]
is independent of $U'$, only the second determinant of $Z_I$
must be expanded around $U'=0$. This yields in leading order
\[
det(-\partial_t+{\bar\mu}-uI+{\hat t})
\approx
\exp[\int\Theta(|\kappa|-1)\log(\kappa)d^dk]
\exp[\beta'h\sum_r(1+S_r)].
\]
with the effective magnetic field
\[
h={u\over2}\int{\Theta(|\kappa|-1)\over \kappa}d^dk
\ \ \
{\rm with}\ \  \kappa={\bar\mu}+2{\hat t}\sum_{j=1}^d\cos k_j.
\]
Thus, the weakly interacting regime prefers a ferromagnetic state in
terms of the Ising spin $S_r=1$.
Knowing the Ising spin configuration with
the highest weight, one can return to the partition function of
Eq. (\ref{partitionf}) and obtains
\[
Z_I\approx [det({\bar\mu}+{\hat t})]^{\beta'}
det(-\partial_t+{\bar\mu}-u+{\hat t}).
\]
The argument of the determinant can be considered as an inverse
Green's function
\[
G=(-\partial_t+{\bar\mu}-u+{\hat t})^{-1}.
\]
Thus, the effect of the ferromagnetic Ising spin structure
is a constant shift of the chemical potential. Of course, the
fluctuations of the Ising spins around the structure with maximal
weight have to be taken into account as soon as $\beta'<\infty$. 

\subsection{Hopping Expansion: ${\bar t}\approx0$}

The expansion of the first determinant of $Z_I$ in powers of $\tau$ is
\begin{equation}
{\bar\mu}^{\beta'|\lambda_k|}
\exp [-{\beta'\over2{\bar\mu}^2}\sum_{r,r\in\lambda_k}
({\hat t}_{r,r'})^2]
\label{weight2}
\end{equation}
and the expansion of the second determinant is
\begin{equation}   
det(-\partial_t+{\bar\mu}-uI+{\hat t})
\approx
{\bar\mu}^{\beta'(M-|\lambda|)}
\exp\Big[
-{\beta'\tau^2\over2}\sum_{\langle r,r\rangle}W(V_r,V_{r'})
\Big].
\label{ising}
\end{equation}
Here the potential terms $V_r,V_{r'}$ are either ${\bar\mu}$ or 
${\bar\mu}-u\approx {\bar\mu}-U'/{\bar\mu}$ with the function
\[
W(V_r,V_{r'})=\cases{
0 & for $V_r=V_{r'}={\bar\mu}-u$ \cr
-1/(u{\bar\mu}) & for $V_r\ne V_{r'}$ \cr
1/{\bar\mu}^2 & for $V_r=V_{r'}={\bar\mu}$ \cr
}.
\]
For $\tau=0$ all terms in the sum $\sum_\lambda Z_\lambda$ have the same
weight.   
This reflects the fact that there is a degenerate perturbation theory
for the hopping expansion in the half-filled system due to the spin
degree of freedom. For $\tau\ne0$ the second factor of (\ref{ising})
can be approximated for $\beta\approx\infty$ ($\beta$ is now real) as
\begin{equation}
\exp\Big[
{\beta'\tau^2\over2}\sum_{\langle r,r\rangle}W(V_r,V_{r'})
\Big]\approx
\exp\Big[-\beta{{\bar t}^2\over2U}\sum_{\langle r,r\rangle}
S_rS_{r'}\Big].
\label{weight3}
\end{equation}
According to (\ref{weight2}) and (\ref{weight3}) the maximal contribution
to $Z_I$ comes from $V_r\ne V_{r'}$ on nearest-neighbor sites.
This corresponds with an antiferromagnet state of the Ising spin
$S_r=(-1)^{r_1+\cdots +r_d}$.
The partition function in Eq. (\ref{partitionf}) reads
with the staggered Ising spin structure
\[
Z_I\approx 2{\bar\mu}^{\beta'M/2}
det(-\partial_t+V_{AFM}+{\hat t}),
\]
where $V_{AFM}$ is the staggered antiferromagnetic potential
\[
V_{AFM}={\bar\mu}-{u\over2}[1+(-1)^{r_1+\cdots +r_d}].
\]
As an effective Green's function of the fermions one can study
\[
G=(-\partial_t+V_{AFM}+{\hat t})^{-1}.
\]
The staggered potential creates a gap. This can be seen, for
instance, in the special case of
a two-dimensional lattice. Then the eigenvalues of $G^{-1}$ are
\[
{\bar\mu} - {u\over2}\pm
\sqrt{{u^2\over4}+|\tau h_{12}|^2}
\]
with $h_{12}=1+e^{ik_x}+e^{ik_y}+e^{ik_x+ik_y}$ ($-\pi\le k_j<\pi$).
This result describes the physics of an insulator with a gap
\[
U/2,
\]
in contrast to the homogeneous potential of the
ferromagnetic Ising structure for the weak interacting case.

There can be a phase transition from weak coupling (ferromagnetic
Ising system) to weak hopping ({\it anti}ferromagnetic Ising system)
even at finite $\beta$ (i.e., non-zero temperature). It is not
clear, however, whether this is a first or second order transition.
Numerical (Monte Carlo) simulations may reveal more details of the
transition.

\section{Conclusions}

Considering quantities of the Hubbard model which depend only
on one spin direction, e.g., spin-down, it is possible to integrate
out the other spin direction in the functional integral. This 
idea was carried out for the partition function $Z$. The latter could
have been extended to a generating function for spin-down Green's
function without modification of the procedure. The result of this 
integration is a functional integral which
has a representation in terms of thermal Ising spins. Moreover,
the fermionic interaction of the effective model has a coupling
parameter $U'{\bar t}$ which becomes small in the limit of weak
Hubbard interaction $U'$ as well as in the limit of low mobility
of the particle (small hopping rate ${\bar t}$). Therefore, the
effective fermionic interaction can be neglected if one is only
interested in the strongly metallic and the strongly insulating
regime at half filling. The advantage of this approach is that the
degeneracy of the special cases $U'=0$ and ${\bar t}=0$, which are
difficult to handle, are controlled by the coupling of $U'$ and
${\bar t}$ to
the Ising spins: a small Hubbard interaction $U'$ creates an effective
magnetic field $h\propto U'$ which couples linearly to the Ising spin.
On the other hand, a small hopping term favours an antiferromagnetic 
(staggered) Ising spin configuration because of the effective
Ising spin interaction
\[
{{\bar t}^2\over2U}\sum_{\langle r,r\rangle}S_rS_{r'}.
\]
The Boltzmann weights of the effective Ising spin system
were used to extract the most relevant contributions to the sum
over the spin configurations. 

The introduction of Ising spins by a direct decoupling of the 
interaction term in Eq. (\ref{action1}) (Hubbard-Stratonovich
transformation) is a well-known approach. It leads to a dynamic
Ising spin \cite{hirsch}. The approach of the present work avoids
the dynamics of the Ising spin. However, the price for this
simplification is a cluster interaction of spinless fermions.

Although the approximations used in this work do not provide insight
into the transition from the metallic to the insulating phase, a
more accurate treatment of the partition function $Z_I$ of
(\ref{partitionf})
by including fluctuations of the Ising spins, e.g., using a Monte Carlo
simulation, may allow to access the transition at not too low
temperatures.

\vskip1cm

\noindent
Acknowledgement  

\noindent
This work was supported by the Sonderforschungsbereich 484.

\section*{Appendix A}

The space-time determinant on the r.h.s. of Eq. (\ref{det})
can also be written as
\[
det\big[-\partial_t+{\bar\mu}+{\hat t}
-U'{\bar\Psi}_\downarrow\Psi_\downarrow\big]=det({\bf 1}+A)
\equiv\sum_{\pi}(-1)^{\pi}
\prod_{(r,t)\in\Lambda}\Big[\delta_{\pi(r,t),(r,t)}+A_{r,t;\pi(r,t)}\Big]
\]
with the matrix $A=-({\bar\mu}+{\hat t}-U'{\bar\Psi}_\downarrow
\Psi_\downarrow)\partial_t^T$.
The product over the lattice sites gives a sum over all subsets
$\Lambda_k\subseteq\Lambda$ of the space $\Lambda$
and their complements $\Lambda_k'=\Lambda\setminus\Lambda_k$
\[
\sum_{\Lambda_k\subseteq\Lambda}\sum_{\pi}(-1)^{\pi}
\Big[\prod_{(r,t)\in\Lambda_k}A_{r,t;\pi(r,t)}\Big]
\Big[\prod_{(r,t)\in \Lambda_k'}\delta_{\pi(r,t),(r,t)}\Big].
\]
The Kronecker delta $\delta_{\pi(r,t),(r,t)}$ on $\Lambda_k'$ implies 
$\pi(r,t)\in \Lambda_k$ for $(r,t)\in \Lambda_k$.
Therefore, only that part of the matrix $A$ contributes 
which is projected onto $\Lambda_k$. 
This implies an expansion of the determinant in terms of all
partitions of the space-time lattice $\Lambda$ as
\begin{equation}
det\big[-\partial_t+{\bar\mu}+{\hat t}
-U'{\bar\Psi}_\downarrow\Psi_\downarrow\big]
=\sum_{\Lambda_k\subseteq\Lambda}det_{\Lambda_k}(P_kAP_k)
\equiv\sum_{\Lambda_k\subseteq\Lambda}det_{\Lambda_k}A
\label{expdet1}
\end{equation}
with $det_{\emptyset}A=1$ for an empty set $\Lambda_k$.
The projected determinant is
\[
det_{\Lambda_k}\Big(-({\bar\mu}+{\hat t}
-U'{\bar\Psi}_\downarrow\Psi_\downarrow)\partial_t^T\Big)
=det_{\Lambda_k}[-({\bar\mu}+{\hat t})\partial_t^T]
det_{\Lambda_k}\Big({\bf 1}-({\bar\mu}+{\hat t})^{-1}
U'{\bar\Psi}_\downarrow\Psi_\downarrow)\Big),
\]
where the second second determinant reads
\[
\exp\Big[
Tr_{\Lambda_k}\log\Big(
{\bf 1}-({\hat t}+{\bar\mu})^{-1}
U'{\bar\Psi}_\downarrow\Psi_\downarrow\Big)
\Big].
\]

\section*{Appendix B}

Expanding the second determinant of $Z_I$ up to first
order in $\delta{\hat V}$ yields
\[
det(-\partial_t+{\bar\mu}+\delta{\hat V}+{\hat t})
\approx det(-\partial_t+{\bar\mu}+{\hat t})
\exp\Big(Tr[\delta{\hat V}(-\partial_t+{\bar\mu}+{\hat t})^{-1}]\Big).
\]
The trace term can be written as
\[
\beta'h\sum_r(1+S_r),
\]
where
\[
h={u\over2}\int{\Theta(|\kappa|-1)\over \kappa}{d^dk\over(2\pi)^d}
\ \ \
{\rm with}\ \  \kappa={\bar\mu}+2\tau\sum_{j=1}^d\cos k_j.
\]

\section*{Appendix C}

The first determinant of $Z_I$ can be expanded up to second
order in ${\hat t}$ as
\[
[det_{\lambda_k}({\bar\mu}+{\hat t})]^{\beta'}
\approx{\bar\mu}^{\beta'|\lambda_k|}
\exp[-{\beta'\over2{\bar\mu}^2}\sum_{r,r'\in\lambda_k}
({\hat t}_{rr'})^2].
\]
The second determinant of $Z_I$ can be expanded in powers of ${\hat t}$
up to second order
\[
det(-\partial_t+{\bar\mu}+\delta{\hat V}+{\hat t})
\]
\[
\approx
det(-\partial_t+{\bar\mu}+\delta{\hat V})
\exp\Big(-{1\over2}Tr[{\hat t}(-\partial_t+{\bar\mu}+\delta{\hat V})^{-1}
{\hat t}(-\partial_t+{\bar\mu}+\delta{\hat V})^{-1}]\Big).
\]
The trace term reads
\[
Tr[{\hat t}(-\partial_t+{\bar\mu})^{-1}
{\hat t}(-\partial_t+{\bar\mu})^{-1}]
=\beta'\sum_{r,r'}({\hat t}_{rr'})^2\int_0^{2\pi}
{1\over V_r-e^{i\omega}}{1\over V_{r'}-e^{i\omega}}{d\omega\over2\pi},
\]
where the integral gives
\[
\int_0^{2\pi}
{1\over V_r-e^{i\omega}}{1\over V_{r'}-e^{i\omega}}
{d\omega\over2\pi}
=\cases{
0 & for $V_r=V_{r'}={\bar\mu}-U'/{\bar\mu}$ \cr
-1/(u{\bar\mu}) & for $V_r\ne V_{r'}$ \cr
1/{\bar\mu}^2 & for $V_r=V_{r'}={\bar\mu}$ \cr
}.
\]
Due to $\tau\approx0$, the approximation $u\approx U'/{\bar\mu}$
can be used. Then for $\beta'\approx\infty$ the determinant is
\[
det(-\partial_t+{\bar\mu}+\delta{\hat V})
=\prod_r(1+[{\bar\mu}+\delta{\hat V}_r]^{\beta'})
\approx {\bar\mu}^{\beta'(M-|\lambda_k|)},
\]
since $|{\bar\mu}|>1$ and $|{\bar\mu}-U'/{\bar\mu}|<1$.

\end{document}